# Entropy, Gravity and the Mass-Boom

Antonio Alfonso-Faus

E.U.I.T. Aeronáutica, Plaza Cardenal Cisneros, 3. 28040 Madrid, Spain

**Abstract. –** Verlinde presents the gravitational force as due to gradients of entropy, an emergent force, with far reaching consequences. Using the Hawking-Bekenstein entropy formulation, we arrive at the conclusion that the Mass-Boom effect, presented elsewhere, forces the entropy of the universe to increase. Then the Mass-Boom is directly related to the existence of gravity. The principle of Mach implies that the Mass-Boom is responsible for the expansion of the universe. Thus, the Mass-Boom effect is a necessary condition for: 1) the increase of entropy with time, 2) the existence of gravity, and 3) for the expansion of the universe. The universe seems to initially appear and grow out of polarization: positive mass-boom (energy) versus negative gravitational potential energy boom, adding both always to zero. Polarization is then the cause of creation and evolution of the universe.

Key words: gravitation, entropy, Mass-Boom, universe, cosmology.

PACS: 95.30Sf, 05.70.-a

## The origin and evolution of the universe

Verlinde´s proposition [1] implies that a non-zero positive gradient of entropy is required for the force of gravity to exist. Taking the gradient in the entropy formulation of Hawking-Bekenstein [2], [3]

$$S = 4\pi\, k/\hbar c\; GM^2 \qquad (1)$$

one has (with k, ђ, c and G constants)

$$\nabla S = 4\pi\, k/ђc\; 2GM\, \nabla M \qquad (2)$$

From (1) the growth of mass, the Mass-Boom [4], implies increasing entropy. Since we need $\nabla S \geq 0$ for the force of gravity to exist, from (2) one must have $\nabla M \geq 0$. The growth of mass is necessary then for the force of gravity to exist. On the other hand Mach´s principle [4] implies that the mass of the universe varies as the cosmological scale factor R(t)

$$GM(t)/c^2 = R(t) \qquad (3)$$

Then, growing mass ensures expansion and from (2) we have a positive gradient of entropy, a gravitational attractive force. The Mass-Boom effect [4] is a necessary condition for: 1) the increase of entropy with time, 2) the existence of gravity, and 3) the expansion of the universe. The existence of the universe appears to be due to the Mass-Boom effect. This is due to a polarization effect: positive mass (energy) versus negative gravitational potential energy always adding to zero:

$$Mc^2 - GM^2/R = 0 \qquad (4)$$

This is the same as the Mach´s principle in (3).